\documentclass[jkps,preprint, fleqn,showpacs,showkeys]{revtex4}
\usepackage{graphicx,amssymb,amsmath,bm}
\begin{document}
\title{Density Functional Theory study on the electronic 
structure and thermoelectric properties of strained Mn$_4$Si$_7$}
\author{Do Duc Cuong}
\author{JinSik Park}
\author{S. H. Rhim}
\email{sonny@ulsan.ac.kr}
\author{Soon Cheol Hong}
\email{schong@ulsan.ac.kr}
\affiliation{Department of Physics and Energy Harvest Storage Research Center,
  University of Ulsan, Ulsan, 680-749, Republic of Korea}
\author{Joo-Hyong Lee}
\affiliation{School of Materials Science and Engineering,
  Gwangju Institute of Science and Technology, Gwangju, Republic of Korea
}

\begin{abstract}
Strain effect on electronic structure and thermoelectric properties of
higher manganese silicides (HMS) Mn$_4$Si$_7$ was studied using density functional
theory (DFT) and through solving Boltzman transport equation (BTE). We found that 
tensile strain reduces band gap while compressive strain
does not much affect the band gap. Seebeck coeficient increases 
with temperature, which is consistent with experiments. The electrical
conductivity and power factor show anisotropy, where in-plane direction is 
more dominant. Anisotropy of the electrical conductivity along in-plane and out
of plane direction was explained due to change of band dispersion in the valence band maximum (VBM).
\end{abstract}



\maketitle

\section{INTRODUCTION}
Search for new source of energy has been rapidly increased recently with high demand 
because of high price and possible exhaustion of traditional energy sources such as fossil fuel, gaseous.
On the other hand, 
the use of current system is not so effective as up to 60\% of energy is wasted as heat 
{\cite {KOUMOTO06}}. In searching the new source of energy, thermoelectric materials receive a lot of interests  
owing to  its capability to convert thermal energy into electrical energy. 
The thermoelectric materials
have many advantages: there is no moving part with high durability, and 
they are environmetally friendly green energy. However, the current state of art of thermoelectrics
is limited in wide-scale applications because of low conversion efficiency and high material costs.
The effciciency of thermoelectric materials is strongly related to dimensionless 
figure of merit, which is defined by $ZT = \frac{\sigma S^2}{\kappa}T$, where $\sigma$, $S$, $\kappa$, 
and $T$ are the electrical conductivity, the Seebeck coefficient, the thermal conductivity, and the absolute temperature,
respectively{\cite {HE13}. Some materials having high ZT such as Bi$_2$Te$_3$, PbTe, and Zn$_4$Sb$_3$, however, contain 
toxic elements or have high cost elements (Te, Sb) {\cite {STEVEN14}}.
Higher Manganese Silides (HMS), MnSi$_{\delta}$, have been considered as alternative promissing thermoelectric materials 
because they exhibit relative high $ZT$, contains two abundant and non-toxic elements.
Furthermore, they possess mechanical stability to corrosion and oxidation {\cite {NISHIDA72,SEARCY57}}.
Depending on $\delta$ value, several closely related phases such as Mn$_4$Si$_7$, Mn$_{11}$Si$_{19}$, and Mn$_{15}$Si$_{26}$,
have been identified in this class of material{\cite {ENGSTROM88, PONNAM11}}.    

There have been a number of studies on thermoelectric properties of HMS. Due to rather low thermal conductivity, 
high Seebeck coefficent, and electrical conductivity, the ZT of HMS is relatively high, whose range is 
0.4-0.75 {\cite {SADIA12,STEVEN14, HOU07}. The improvement of 
the ZT in the HMS seems fascinating. In this study, the strain effect on the electronic 
structure and thermoelectric properties of HMS has been investigated
using first-principles calculations and Boltzman Transport equation (BTE).
\section{METHODOLOGY}
First-principles density functional calculations are performed using 
Vienna {\em Ab Initio} Simulation Package (VASP)\cite{VASP}.
For the exchange-correlation potential,
generalized gradient approximation (GGA) is employed
with Perdew, Burke, and Ernzerhof (PBE) parametrization\cite{GGA,PBE}
with projected augmented wave (PAW) basis sets\cite{PAW}. 
Cutoffs for kinetic energy is chosen 400 eV. 
The Brillouin-zone (BZ) integration is done with 6$\times$6$\times$2 {\em k} mesh
 in Monkhorst-Pack scheme.
BoltzTrap\cite{BOLTZTRAP} is used 
interpolate band eigenvalues to extract velocities,
which is based on the converged electronic structure.
The thermoelectric properties are calculated based on the interpolation
on {\em k} grid of 12$\times$12$\times$4 giving 144 {\em k} point
in the irreducible Brillouin zone wedge.
Throughout this paper, we consider Mn$_4$Si$_7$, where $\delta$=1.75.
\section{RESULTS AND DISCUSSION}
Mn$_4$Si$_7$ is well-known as a degenerate $p-$type semiconductor with hole carrier concentration of order of $10^{21}cm^{-3}$. 
Crystal structure belongs to the well-known Nowotny Chimney-Ladder compounds, consisting of two 
sublattices : Manganese "chimney" sublattice and Silicon "ladder" sublattice {\cite {MIYAZAKI08}}.
Calculated lattice constants are $a$ = 5.508 $\AA$ and $c/a$ = 3.158, which agrees well with experiments and 
other previous calculations {\cite {KAJITANI10, MIYAZAKI08,MIGAS08}}. The electronic band structure of Mn$_4$Si$_7$ show 
semiconducting behavior with indirect band gap of 0.82 eV between $\Gamma$-points in 
VCBM and $Z$-point in CBM, also consistent with previous calculations {\cite {CAPRARA08,MIGAS08}}

To study the strain effect, the different strain is applied from -3\% (compressive strain) to +3\% 
(tensile strain) along $x$-axis ($a$) and $y$-axis ($b$) respect to the unstrained lattice.
Lattice constant along $z$-axis ($c$) is optimized for each set of lattice constant along $x$- 
and $y$-axis. As a result, change of $c$ is from 5.1\% to -4.4\%, compared to unstrained lattice, 
This change of $c$ leads to the increase of volume from -4.2\% to +4.4\% as shown in Fig.~\ref{fig1}(a).

The change in band gap is more dramatic as shown in Fig.~\ref{fig1}(b). The band gap is not much changed 
under compressive strain, but does change under the tensile strain. The band gap 
is reduced from 0.82 eV to 0.77 eV at +3\% tensile strain with respect to the unstrained one. 
This trend in band gap change is totally different from other materials such as SnSe, 
where the band gap was found to increase due to the increasing of lattice 
separation along $z$ direction \cite {CUONG15}. 
\begin{figure}
  \centering
  \includegraphics[width=\columnwidth]{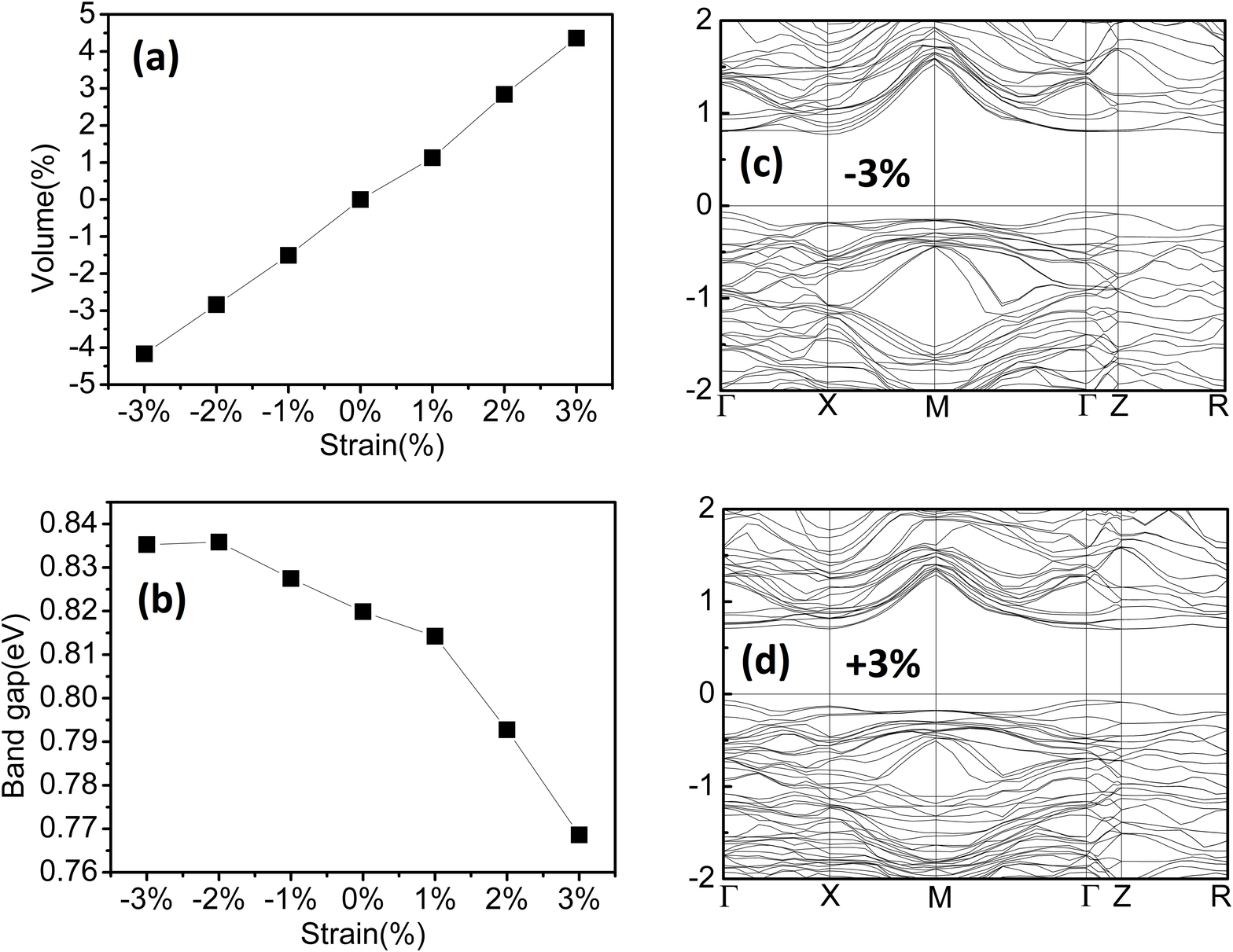}
  \caption{Dependence of volume (a) and band gap (b)
    with different strain from -3\% to +3\% and electronic band structure of Mn$_4$Si$_7$ 
    in -3\% compressive strain (c) and +3\% tensile strain (d).
  }
  \label{fig1}
\end{figure}

In Fig.~\ref{fig1}(c) and (d), we show the comparison of band structure of compressive (-3\%) 
and tensile strain (+3\%) case. Other strains are not shown here but the general trends are similar. 
The band structure topology is alsmost same under applying strain, except for band gap. 
This behavior can be predictable due to the fact that the properties of HMS Mn$_4$Si$_7$ is stable
 under ambient conditions as reported in previous experiments {\cite {NISHIDA72,SEARCY57}}. The band dispersion
on VBM along the $\Gamma-Z$ direction is quite flat compared to along the $\Gamma-X$ direction. 
However, under
applying compressive strain, the band along  $\Gamma-Z$ on VBM is slightly 
more dispersive, while it is less dispersive along $\Gamma-X$ direction. 
This change of band dispersion will affect the electrical 
conductivity which will be discussed later.  

The Seebeck coefficient, electrical conductivity $\sigma/\tau$, and 
 power factor $PF/\tau = \sigma/\tau.S^2$ were calculated and showed in 
Fig.~\ref{fig2} as the function of temperature, where $\tau$ is relaxation time.
 In all calculations, the hole carrier concentration was 
fixed at experimental value $10^{21} cm^{-3}$ \cite {SADIA12}. 
The Seebeck coefficient in unstrained Mn$_4$Si$_7$ increases with increasing 
temperature, which show very good agreement with experimental data \cite {SADIA12},
However, the electrical conductivity $\sigma/\tau$ show highly anisotropic, 
where the $\sigma/\tau$ along the in-plane direction is almost three times bigger compared with that 
along out of plane direction. The large difference between in-plane and out of plane electrical 
conductivity come from the difference in band dispersion on the VBM which we discussed above, because 
the electrical conductivity is proportional to the electron verlocity, which is defined as the derivate of 
energy band versus the k-vector \cite {BOLTZTRAP}. As a result, the $PF/\tau$  along in-plane lattice 
is much higher compared with along out of plane lattice. 

\begin{figure*}
  \centering
\includegraphics[width=\columnwidth]{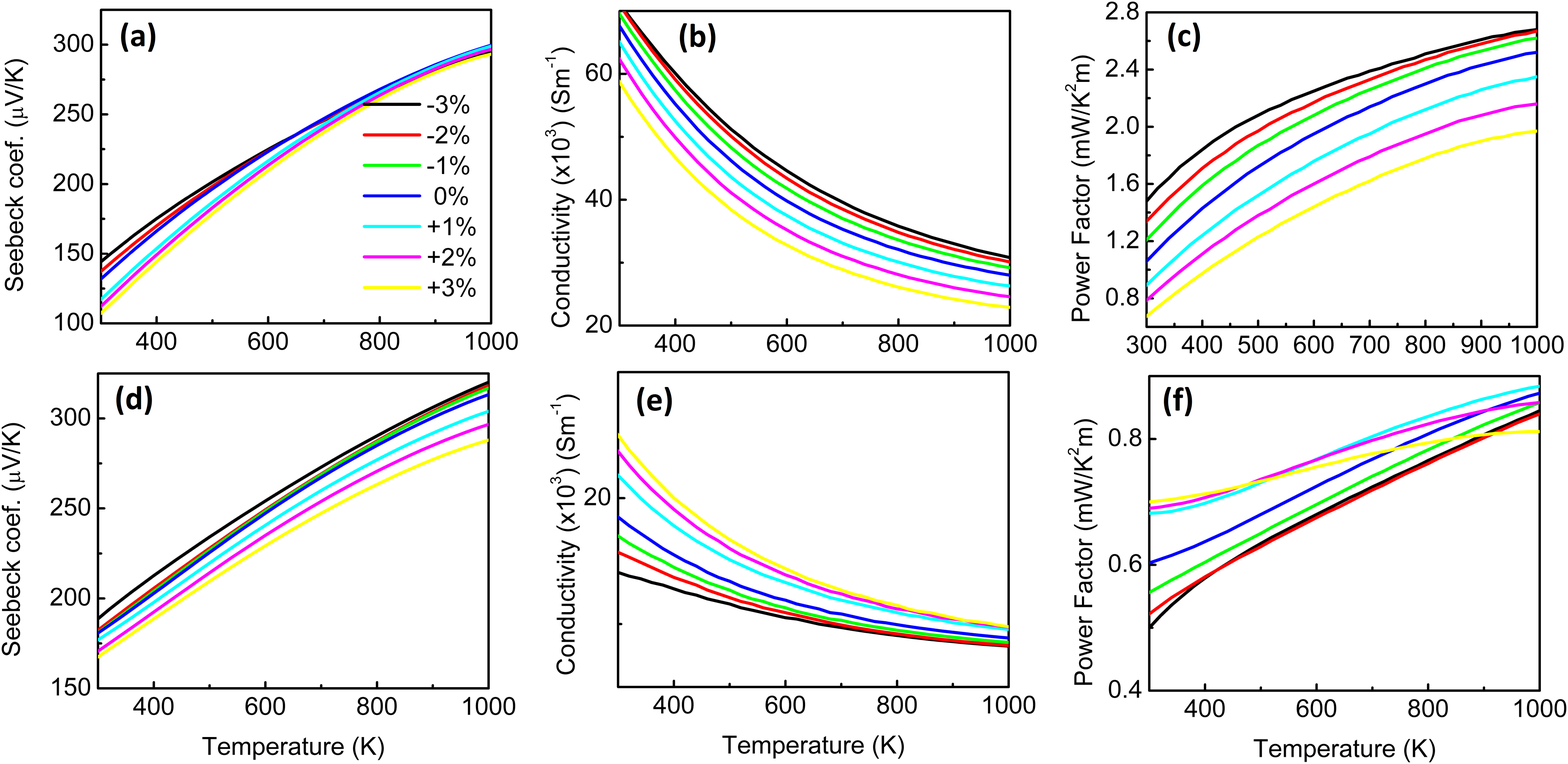}
\caption{
  The effect of strain on the thermoelectric properties of Mn$_4$Si$_7$: 
  The Seebeck coefficient (a), electrical conducivity $\sigma/\tau$ (b) and Power factor (c) along 
  in-plane direction. 
  The Seebeck coefficient (d), electrical conductivity $\sigma/\tau$ (e) and Power factor (f) along 
  out of plane direction. 
}
  \label{fig2}
\end{figure*}

Under the strain, the Seebeck coefficient does not change very much 
in both in-plane and out of plane direction. 
However, the behavior of electrical conductivity is 
quite different and exhibits some anisotropy.
Due to the change of band dispersion in the VBM, the $\sigma/\tau$ show different behavior, 
$\sigma/\tau$ along in-plane direction decreases under the tensile strain,
and increase under the compressive strain.
While, $\sigma/\tau$ along out-of-plane direction is reduced under the compressive strain and 
increases under the tensile strain. As a results, the power factor (PF) show the same trend 
with the electrical conductivity. It is noticeable that, even reduction in the out-of-plane lattice, 
the PF along the in-plane lattice is more dominant.
Therefore, the increase of PF along this direction leads to increase the overall PF and ZT. 
\section {CONCLUSION}
The strain effect on the electronic structure and thermoelectric properties of Mn$_4$Si$_7$ has been studied
using Density functional theory calculations and solving Boltzman Transport Equation.
The Seebeck coefficient increases with temperature.
Both the electrical conductivity and PF increase along the in-plane 
direction, and decrease along out-of-plane direction as a result of band dispersion change at the VBM.
We also suggest that the compressive strain can improve PF and ZT of Mn$_4$Si$_7$.
\section {ACKNOWLEDGEMENT}
This work is supported by a grant from
Energy Efficiency \& Resource program of the Korea Institute of Energy Technology Evaluation and Planning 
(KETEP) funded by the Korean Ministry of Knowledge Economy (20132020000110),
and Priority Research Centers Program (2009-0093818) through the NRF funded by the Ministry of Education.



\end{document}